\documentclass[aps,twocolumn,superscriptaddress,nofootinbib,amsmath,amssymb,floatfix,multicol]{revtex4}  
\usepackage{graphicx,caption}  
\usepackage{dcolumn}   
\usepackage{bm,color}        
\usepackage{amssymb}   

\usepackage{adjustbox}
\usepackage{hyperref}
\DeclareMathOperator{\Tr}{Tr}

\newcommand{\be}{\begin{equation}}
\newcommand{\ee}{\end{equation}}
\newcommand{\beq}{\begin{equation}}
\newcommand{\eeq}{\end{equation}}
\newcommand{\bea}{\begin{eqnarray}}
\newcommand{\eea}{\end{eqnarray}}

\newcommand{\aT}{\alpha_{\rm T}}
\newcommand{\aB}{\alpha_{\rm B}}
\newcommand{\aM}{\alpha_{\rm M}}
\newcommand{\aK}{\alpha_{\rm K}}
\newcommand{\aH}{\alpha_{\rm H}}
\newcommand{\aX}{\alpha_{\rm X}}
\newcommand{\sei}{\sigma_8}

\definecolor{MyBlue}{rgb}{0.15,0.15,0.70}

\hypersetup{
colorlinks=true,
citecolor=red,
linkcolor=MyBlue,
urlcolor=MyBlue
}

\begin{document}
\title{Parametrizing modified gravity for cosmological surveys}
\author{J\'er\^ome Gleyzes}

\address{Jet Propulsion Laboratory, California Institute of Technology, Pasadena, CA 91109
         California Institute of Technology, Pasadena, CA 91125}

        \date{\today}

\begin{abstract}
One of the challenges in testing gravity with cosmology is the vast freedom opened when extending General Relativity. For linear perturbations, one solution consists in using the Effective Field Theory of Dark Energy (EFT of DE). Even then, the theory space is described in terms of a handful of free functions of time. This needs to be reduced to a finite number of parameters to be practical for cosmological surveys. We explore in this article how well simple parametrizations, with a small number of parameters, can fit observables computed from complex theories. Imposing the stability of linear perturbations appreciably reduces the theory space we explore. We find that observables are not extremely sensitive to short time-scale variations and that simple, smooth parametrizations are usually sufficient to describe this theory space. Using the Bayesian Information Criterion, we find that using two parameters for each function (an amplitude and a power law index) is preferred over complex models for $86\%$ of our theory space.

\end{abstract}

\maketitle

\section{Introduction}
With many large scale structure surveys, such as WFIRST \cite{Spergel:2013tha}, LSST \cite{Abell:2009aa}, DESI \cite{Aghamousa:2016zmz}, SPHEREx \cite{spherex_wp} and EUCLID \cite{Laureijs:2011gra} coming online in the close future, our chances of understanding what is causing the accelerated expansion of the Universe are improving drastically. While a cosmological constant is still consistent with the data, it is informative to see how deviations from our standard cosmological model $\Lambda$CDM could be constrained with these upcoming experiments.

It was shown in the broad framework of the Effective Field Theory of Dark Energy (EFT of DE) \cite{Gubitosi:2012hu,Bloomfield:2012ff,Gleyzes:2013ooa} that to describe modifications of gravity involving a single extra degree of freedom (DOF), such as a scalar field, only five free functions of time are needed\footnote{More functions are needed if the weak equivalence principle is broken, i.e.~dark matter and baryons are coupled to different metrics \cite{Gleyzes:2015pma}.}. In particular, they can be used to describe known models such as Horndeski theories \cite{Horndeski:1974wa} and their extensions \cite{Zumalacarregui:2013pma,Gleyzes:2014dya,Gleyzes:2014qga}. The case of Horndeski only requires four out of the five functions of time, and they have been expressed in a insightful and convenient way in \cite{Bellini:2014fua}. The addition of beyond Horndeski  theories to the notation of \cite{Bellini:2014fua} was presented in \cite{Gleyzes:2014rba}. Note that recently, the formalism was extended to include also the possibility of modifications of gravity including additional vectors and tensors\cite{Lagos:2016wyv,Lagos:2016gep}.

When used in the context of a specific model, these functions of time are not free, but can be computed once the parameters of said model are given. Doing so requires solving the background equations to get the associated time evolution, so that in principle the EFT of DE can be used as a proxy for specific models.  However, no clear candidate stands out as a promising alternative to General Relativity (GR), which means that one should probably not focus only on these models.

The EFT of DE then becomes particularly critical: there is no need to specify a given model, since the appearance of these five functions of time arises generically when allowing the presence of a scalar field. The framework allows to thoroughly and systematically explore the theory space around $\Lambda$CDM, and let observations highlight which regions of this theory space are the most favored. The efforts for building models could then be focused to these particular regions. The difficult part there is that one has to deal with free functions of time, which are difficult to constrain with the limited observations that we have.

\begin{figure}[h!]
\centering
\includegraphics[width=0.5\textwidth]{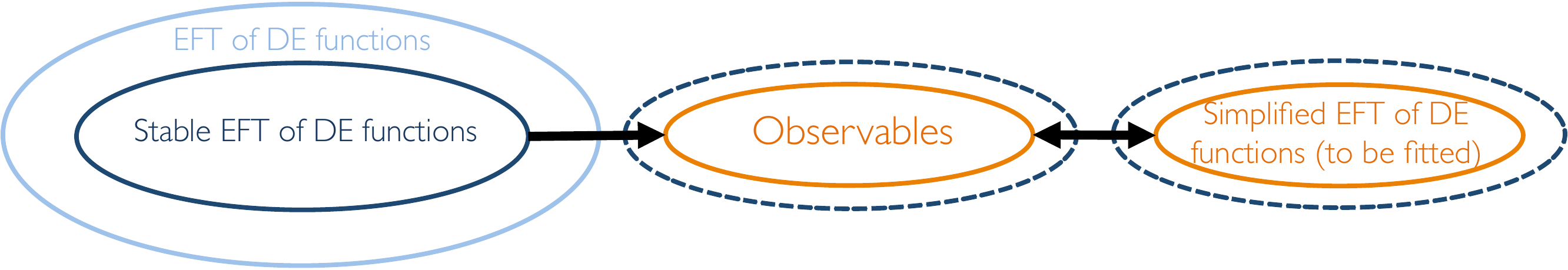}
\caption{Representation of the method used in this paper. We will compute observables (galaxy and weak lensing power spectra) from complex EFT of DE functions after imposing stability conditions. They will be fitted to simplified versions of these functions. Then we will use model comparison tools to assess whether the complexity (i.e.~the size of the circle) of the original functions is actually transferred to the observables, or if simple functions are enough capture the physical features. In this diagram, this essentially corresponds to comparing the size of the dark blue and the orange circles. Moreover, we want the orange circle on the right to be the same size as the middle one, i.e.~the simple functions have to explain the whole observable space.}
\label{fig:Diagram}
\end{figure}

The goal of this paper is to try and see if, given the sensitivity of future surveys, one can approximate the complicated landscape of the arbitrary time dependences by a much simpler theory space, only given by a few parameters. We summarize our method in fig.~\ref{fig:Diagram}.
\vspace{-0.2cm}
\vspace{-0.2cm}

In Section \ref{sec:timdep}, we briefly review and expose the features of the EFT of DE that are relevant for our analysis. Then, in Section \ref{sec:setup}, we present the way we are going to explore the theory space, and how we are going to estimate the performance of simpler parametrizations. The results are then detailed in Section \ref{sec:results}, and final conclusions are drawn in Section \ref{sec:ccl}.
\vspace{-0.35cm}
\section{The Framework}
\label{sec:timdep}
The EFT of DE formalism has been carefully reviewed in \cite{Gleyzes:2014rba}. For our purpose, the key conclusion is that assuming the background is fixed, the evolution of perturbations is in principle determined by five free functions of time. In Ref.~\cite{Bellini:2014fua} and later \cite{Gleyzes:2014rba}, those functions have been expressed in a way that highlights their effects on the theory space, and are written as $\{\aK,\aB,\aT,\aM,\aH\}$. 

\begin{itemize}
\item $\aK$ parametrizes the kinetic energy of the extra degree of freedom and is the simplest extension to GR. Setting all the other functions to zero will capture simple dark energy models (i.e.~with no modifications to gravity except the presence of an extra fluid). 
\item$\aB$ is linked to so-called braiding scalar-tensor models \cite{Deffayet:2010qz} where part of the kinetic energy of the scalar is sourced through a coupling to gravity, resulting in deeper modifications of gravity. 
\item $\aT$ controls the deviation of the speed of tensor perturbations from that of light, which is allowed when going outside of GR. It could in principle be measured independently using gravitational waves \cite{Bettoni:2016mij}.
\item $\aM$ is non-zero when the Planck mass, denoted $M(t)$ and defined as the coupling between gravity waves and matter, is not constant in time. In scalar-tensor theories, this happens when the scalar field couples directly to the Ricci scalar.
\item $\aH$ vanishes for theories that belong to the Horndeski class \cite{Horndeski:1974wa}, where terms with more than two derivatives are forbidden from the equations of motion. However, a proper interpretation of Ostrogradski's theorem \cite{Woodard:2006nt} indicates that higher derivatives are not necessarily synonym of instabilities. What is important is that the actual degrees of freedom, once constraint equations are solved, obey second order differential equations.  This is the case in theories dubbed beyond Horndeski \cite{Zumalacarregui:2013pma,Gleyzes:2014dya,Gleyzes:2014qga} (see also \cite{Deffayet:2015qwa,Langlois:2015cwa,Langlois:2015skt,Achour:2016rkg,Crisostomi:2016tcp}), and leads to $\aH\neq 0$. 
\end{itemize} 
The advantage of using this set of functions (instead of the original set of \cite{Gubitosi:2012hu,Bloomfield:2012ff,Gleyzes:2013ooa}) is that, if any of them were measured to be non-zero, it would directly point towards one specific aspect of modified gravity. As a shorthand notation, we will denote this set by $\{\aX\}$.

Using these functions, one can derive the modified Einstein equations as well as an evolution equation for the extra degree of freedom. If one further supplements them with the conservation of the matter stress energy tensor, the system can be solved to obtain the evolution of a (linear) matter overdensity. However, even if the functions $\{\aX\}$ are taken to have simple and known time dependences, the system of equations can only be solved numerically. Different groups have developed codes in order to do so, such as EFTCAMB \cite{Hu:2013twa}, Hi-Class \cite{Zumalacarregui:2016pph} or COOP \cite{DAmico:2016ntq}.
Here, for the sake of simplicity, we will look at the simpler case of the extreme quasi-static (EQS) limit, that is justified when looking at overdensities on scale much smaller than the sound horizon of the extra DOF \cite{Sawicki:2015zya}. To make this more quantitative, let us explicitly introduce the metric in Newtonian gauge (which is convenient for phenomenological studies). The line element reads
\be
{\rm d}s^2=-(1+2\Phi){\rm d}t^2+a(t)^2(1-2\Psi){\rm d}\vec x^2\,.
\ee
Since this analysis is limited to linear perturbations, it is easier to go to Fourier space. Taking the EQS means looking at wavenumbers $k\gg c_s^{-1} aH $, where $c_s$ is the speed of sound of the extra DOF (that can be expressed in terms of the $\{\aX\}$), and the Hubble rate is defined as $H\equiv \dot a/a$. We will also limit ourselves to the case $\aH=0$ to keep expressions simpler. In this limit, one recovers a growth equation for matter overdensities $\delta_{\rm m}$ that is similar to that of GR, given by (omitting the explicit time dependences for brievity)
\begin{align}
\ddot \delta_{\rm m}+2H\dot \delta_{\rm m}&=\frac32\Omega_{\rm m} H^2\mu_{\rm eff}\delta_{\rm m}\,, \label{eq:growth}\\
\mu_{\rm eff}\equiv&\, (1+\aT+\beta_\xi^2)\,,\label{geff}
\end{align}
where we have defined 
\be
\Omega_{\rm m}\equiv \frac{\rho_{\rm m}}{3H^2M^2}\,,\quad \beta_\xi\equiv  \frac{\sqrt{2} }{  c_s \alpha^{1/2} }\left[ \aB (1+\aT) + \aT - \aM\right]\,.
\ee
$\alpha\equiv \aK+6\aB^2$ is the total kinetic energy of the scalar field, and its speed of sound is given as
\begin{align}
c_s^2 \equiv -\frac{2}{\alpha}\bigg\{ &(1+\aB) \bigg[ \frac{\dot H}{H^2} -\aM + \aT  +  \aB (1+\aT)\bigg] \nonumber\\
&+ \frac{\dot \alpha_{\rm B}}{H} +\frac32 \Omega_{\rm m} \bigg\} \,.\label{eqcs}
\end{align}
Stability conditions impose that both $\alpha>0$ (no-ghost condition, see \cite{Sbisa:2014pzo}) and $c_s^2>0$ (no gradient instability),  ensuring that $\beta_\xi$ is real. Even in the absence of any anisotropic stress, modified gravity models do not have $\Phi=\Psi$ as in GR. This is important, as weak lensing measurements are sensitive to the so-called lensing potential, $\Psi+\Phi$. In terms of the $\{\aX\}$, this potential can be written as
 \begin{align}
\label{LensingPot}
  &\Phi + \Psi=-\frac{3a^2H^2}{k^2}\Omega_{\rm m}\mu_{\rm light}\delta_{\rm m}\, ,\\
  \mu_{\rm light}\equiv &\,\frac{2+\aT+(\beta_\xi+\beta_B) \beta_\xi}2\,,\quad \beta_B\equiv \frac{\sqrt{2}\aB }{  c_s \alpha^{1/2} }\,.\label{defGl}
\end{align}
If $\aT=\aM=0$, then one finds that $\Phi=\Psi$, but otherwise, they are different.

Thus, to get the power spectrum at a given redshift, one needs in principle the whole evolution of the $\{\aX\}$. If we had perfect knowledge of the evolution of the matter field from high redshifts to today, one could create a fine binning in redshift and associate a parameter for each functions in each bins. For realistic surveys however, it would be better if those free functions of time could be fixed in terms of a few constant parameters. This is why a number of groups \cite{Bellini:2014fua,Perenon:2015sla,Gleyzes:2015rua,Alonso:2016suf,Casas:2017eob} have decided, as a first attempt, to parametrize the functions of time as proportional to $1-\Omega_{\rm m}$. Indeed, under the assumption of a spatially flat Universe, $\Omega_{\rm m}=1$ if the Universe contains only matter, and is smaller if there are other components. Therefore, $1-\Omega_{\rm m}$ controls the density of what is causing the accelerated expansion of the Universe\footnote{The density of radiation is negligible in the recent Universe, so that it can safely be ignored from $1-\Omega_{\rm m}$.}. This is why it is pretty natural to assume that deviations to $\Lambda$CDM would trace this quantity.

Of course, the time evolution could be a more complicated function of $1-\Omega_{\rm m}$, and it could be different for every function $\{\aX\}$. There have been attempts to see whether this simple parametrization reproduces known models, such as galileons \cite{Deffayet:2009wt}, with some indicating reasonable agreement \cite{Renk:2016olm} while other claiming the opposite \cite{Linder:2016wqw}. Since it is not clear whether they probed the same part of the theory space, this is not necessarily a contradiction. However, galileons are not necessarily representative of the full theory space that we are trying to explore. Nor are they the most motivated candidates from a cosmological point of view, having trouble with observations due to an unusual sign for the ISW effect \cite{Barreira:2014jha}. Therefore, even if simple parametrizations turned out to not be working very well for these models, that would not undoubtedly mean that they should be thrown away.

In this paper, we argue that whether a simple parametrization can reproduce the evolution of the $\{\aX\}$ coming from complicated models is not the most relevant question. Indeed, it is not obvious the complicated features in the  $\{\aX\}$ are actually observable. For instance, their effect on the growth is integrated, as seen in eq.~\eqref{eq:growth}, so one might guess that sharp features are smoothed out.
Instead, we would rather see whether the difference between simple and complicated evolution of the $\{\aX\}$ can actually be measured.

 In order to try and answer this question, we will do the following: take the $\{\aX\}$ to be random, different functions of time, and compute the corresponding power spectra, for galaxies and weak lensing. Then, we will try and see if we can fit the observables using the simple parametrizations and minimizing the $\chi^2$ for said observables.
This differs from previous analyses in three major ways: because we take the functions to be random, the exploration of the theory space is not biased towards a specific model. Secondly, for each value of the $\{\aX\}$ today, we will take many different realizations of the random functions, so that we can make quantitative statement about how often the parametrization fails. Finally, the comparison and fitting will be at the level of the observables, not of the functions $\{\aX\}$ nor $\mu_{\rm eff,light}$, which are not directly measurable.
 
 \section{The setup}
 \label{sec:setup}
We want to explore the theory space of the $\{\aX\}$ and see if it can be approximated by a simple, finite dimensional parameter space. To do so, we can start from a very large space, essentially mimicking the infinite dimensional space, and see if the simple parametrization allows to recover the features of this complex parameter space that are relevant for cosmological surveys, i.e.~those conveyed to the observables.

Before detailing the exact procedure, let us note that within the approximations that we have described in the previous section, particularly the EQS, only three functions $\{\aB,\aT,\aM\}$ have an effect. We set $\aH=0$ and $\aK$ does not appear in the equations. For each of the three remaining functions, we will parametrize the ``true" theory space as
\be
\label{truealpha}
\alpha^{\rm  true}_{\rm X}(z)=\alpha_{{\rm X},0}(1+z)^{-q_{\rm X}}\left(\frac{1+\sum_{i=1}^{i_{\rm max}} n_{{\rm X},i} z^i}{1+\sum_{i=1}^{i_{\rm max}} d_{{\rm X},i} z^i}\right)\,,
\ee
where ${\rm X}\in\{\rm B,T,M\}$. We express the time variable as the redshift $z=a(t)^{-1}-1$, and $\alpha_{{\rm X},0}$ is the value of $\aX$ today ($z=0$). The choice of the factor $(1+z)^{-q_{\rm X}}$ is to ensure that $\aX$ goes to zero in the past, because we want the effect of modified gravity to only become manifest in the late universe, not during matter domination ($z\gg1$). This is also a generic feature of Horndeski models, as pointed out in \cite{Linder:2016wqw}. Finally, the last part of this function allows for complicated evolutions. The choice of a rational function and not of a polynomial is because typically, the $\{\aX\}$ are defined as ratios of functions that involve the extra degree of freedom (see e.g. \cite{Bellini:2014fua}).

We have checked that this form can fit the case of $k$-essence \cite{ArmendarizPicon:2000dh}, where only $\aK$ is non zero, and goes as $1-\Omega_{\rm m}$ \cite{Bellini:2014fua}. It also works for more involved cases, such as galileon models \cite{Appleby:2011aa}. To check that, we numerically solved the full background equations\footnote{Using a python script generously provided by Alexandre Barreira.}, computed the $\{\aX\}$ and fitted eq.~\eqref{truealpha} to them. We found that with $i_{\rm max}=8$, one could fit the $\{\aX\}$ from the galileon models to better than $10^{-5}$ accuracy. We will thus assume this value from now on. 

The procedure is then as follow: we first choose a triplet $\{\alpha_{{\rm B},0},\alpha_{{\rm T},0},\alpha_{{\rm M},0}\}$ for $\alpha_{{\rm X},0} \in [-1,1]$. This is a way to enforce an observation prior, excluding models potentially ruled out by current observations. For each triplet, we choose at random the three $\{q_{\rm X}\}$, between $[2,6]$, and the $6\,\times\, i_{\rm max}$ parameters $\{d_{{\rm X},i},n_{{\rm X},i}\}$ between $[-100,100]$. If the corresponding $c_s^2 \alpha$ computed from eq.~\eqref{eqcs} crosses zero anytime between $z=0$ and $z=z_{\rm ini}=20$, we discard this realization. Otherwise, we compute the evolution of $\delta_{\rm m}$, setting the initial conditions in matter dominance ($z=z_{\rm ini}$).  Since we choose our functions to decay in principle at least as $(1+z)^{-2}$ for $z\gg1$, their effect is negligible and we have the usual solution $\delta_{\rm m}(z\gg1)=(1+z)^{-1}$. We summarize the choice of priors in Table \ref{tab:prior}.
\renewcommand{\arraystretch}{1.4}
\begin{table}[h]
\small
\begin{center}
\begin{adjustbox}{max width=\textwidth}
\begin{tabular}{|c|c|c|c|}
  \hline
 $\alpha_{\rm X,0}$ &  $q_{\rm X}$ &    $\{d_{{\rm X},i},\, n_{{\rm X},i} \}$ &  Theory priors   \\
  \hline
 $[-1,1]$ &  $[2,6]$ &  $  [-100,100] $ &$c_s^2\alpha>0$ and  \\
& & &  $|(\mu_{\rm light})\sigma_8/\sigma_8^{\Lambda{\rm CDM}}-1|<0.25$ \\
 \hline
  \end{tabular}
\end{adjustbox}
\end{center}
\caption{ The choice of (linear) priors on the parameters of eq.~\eqref{truealpha}. For the theory priors, we impose that both $\sigma_8$ and $\mu_{\rm light}\sigma_8$ are within $25\%$ of the $\Lambda$CDM value.}
\label{tab:prior}
\vspace{-0.2cm}
\end{table}

The final cut that we make on the theory space is that, for each redshift bin, we compute $\sei$ for the matter power spectrum (its amplitude in a sphere of radius $8h$/Mpc) and $\mu_{\rm light}\sei$, the one associated to weak lensing. The current errors on $\sei$ from redshift-space distortions and weak lensing are of order $15\%$ \cite{Macaulay:2013swa,Kilbinger:2014cea}. Therefore, if the $\sei$ and $\mu_{\rm light}\sei$ that we find are not within $25\%$ (to be conservative) of those computed in $\Lambda$CDM, we reject the realization. This is to be consistent with the fact that no deviation from $\Lambda$CDM has been observed.
Then, we want to know if the observables produced with the complicated $\{\aX\}$ in eq.~\eqref{truealpha} can be fitted with simple parametrizations. Within the representation of fig.~\ref{fig:Diagram}, this means checking that the two orange circles are the same size. In order to do this, we will fit the true model to given parametrizations by minimizing the $\chi^2$ for the combination of two probes, galaxy clustering and weak lensing tomography, and then compute the associated Bayesian Information Criterion (BIC), defined as
\be
{\rm BIC}\equiv \chi^2_{\rm min}+k \ln N\,,\label{defBIC}
\ee
with $k$ the number of parameters to be fitted, and $N$ the number of data points. The last term is to penalize models with unnecessary complexity.

\subsection{Simple parametrizations}
\label{sec:params}
The procedure described above will be applied to different parametrizations, with increasing level of complexity. At the end, they will be compared to the ``true" model, as well as each other.

\begin{itemize}
\item With a $w$CDM model (all the $\{\aX\}=0$), and $w\in[-1.02,-0.92]$, using prior knowledge from distance measurements \cite{Aubourg:2014yra}. This is to check whether one really needs modified gravity, or if dark energy could explain the observables.
\item With a parametrization given by
\be
\aX=\alpha_{{\rm X},0}\,\frac{1-\Omega_{\rm m}}{1-\Omega_{\rm m,0}}\,, \label{paramOm}
\ee
where $\Omega_{\rm m,0}$ is the current value of the density parameter $\Omega_{\rm m}$. We will label it ``$\Omega_{\rm m}$". This one has been commonly used in forecasting papers \cite{Bellini:2014fua,Perenon:2015sla,Gleyzes:2015rua,Alonso:2016suf,Casas:2017eob}. Note that there are two assumptions behind it: all of the  $\{\aX\}$ have the same time dependence, and that time dependence is fixed.
\item With a parametrization given by 
\be
\aX=\alpha_{\rm X,0}(1+z)^{-q}\,,\label{param1i}
\ee
with the same $q$ for all the $\{\aX\}$. We will label it ``1i'', for one index. Here, we relax the assumption of a fixed time dependence, but keep all of the $\{\aX\}$ proportional to the same function.
\item With a parametrization given by
\be
\aX=\alpha_{\rm X,0}(1+z)^{-q_{\rm X}}\,,\label{param3i}
\ee
with a different $q_{\rm X}$ for each $\aX$. We will label it ``3i'', for three indices. We have relaxed the two assumptions of the case ``$\Omega_{\rm m}$". We only impose that the functions have this simple redshift dependence, which gets negligible at high redshifts.
\end{itemize}
In every parametrization, the background is a $w$CDM, with $w$ determined by the fitting procedure. More precisely, its range $w$ is increased to $[-1.2,0.8]$. This is to be conservative, since in principle, a non-zero $\alpha_{\rm M}$ (i.e.~a time varying Planck mass) changes the expansion history, inducing degeneracies in the determination of $w$ from distance measurements. Finally, the parameters $\{\alpha_{\rm X,0}\}$ are allowed to vary between $[-50,50]$ and the indices $\{q_{\rm X}\}$ between $[0,6]$.

In order to compare with the ``true" model, we will also compute the BIC in the case where one would fit eq.~\eqref{truealpha} to the galaxy and weak lensing power spectra. This corresponds to having a perfect fit $\chi^2=0$, but it is penalized because of the high number of parameters (18 for each function $\{\aX\}$ plus $w$). By comparing the BIC, we want to see whether the data requires such a high number of parameters to describe the observables. 

Note that in principle, even if the Universe were given by eq.~\eqref{truealpha}, the measurements would have noise, which means that the $\chi^2$ computed with the ``true" model would not be zero (it would roughly be of order the number of measurements see e.g.~\cite{Verde:2009tu}). However since here we want to look at the most difficult scenario to test the approximations, we deliberately choose not to include noise, to see if even then, one can use simple parametrizations for data analyses.
\vspace{-0.5cm}
\subsection{Galaxy clustering}

For galaxy clustering, we assume a spectroscopic redshift survey of $15 \, 000$ squared degrees, sliced in eight equally-populated redshift bins (we take the galaxy distribution as given by \cite{Geach:2009tm} with a limiting flux  placed at $4 \times 10^{-16} \, \text{erg} \, \text{s}^{-1} \, \text{cm}^{-2}$) between $z=0.5$ and $z=2.1$, as in \cite{Gleyzes:2015rua}. These characteristics are similar to those expected in DESI \cite{Aghamousa:2016zmz} or EUCLID \cite{Laureijs:2011gra}. We then compute the $\chi^2$ for a model $\mathcal{M}$
\be\label{eq:like}
\begin{split}
\chi^2_{{\rm PS},\mathcal{M}}\equiv&\sum_{ k,i} \left[P_{\rm true}(k,z_i)-P_{\mathcal{M}}(k,z_i)\right]^2\sigma_{k,i}^{-2}\,,\\
 &\sigma_{k,i}^2\equiv N_{k,i}^{-1}\,P_{\rm true}(k,z_i)^2 \,,
\end{split}
\ee
where $N_{k,i}\equiv \frac{k^2 V_i}{2\pi^2}\Delta k$ is the number of modes in a $k$-bin $[k,k+\Delta k]$ for a redshift bin centered around $z_i$, whose volume is $V_i=V(z_i)$. $P_{\rm true}(k,z)$ is the power spectrum at redshift $z$ and wavenumber $k$, computed with a given realization of eq.~\eqref{truealpha}, and $P_{\mathcal{M}}$ is the one computed with a model $\mathcal{M}$ where the $\{\aX\}$ are given by $w$CDM or either one of eqs.~\eqref{paramOm}--\eqref{param3i}. 

In the EQS approximation, the $k$ dependence of the power spectrum is not modified by the deviations to $\Lambda$CDM. Thus, it drops out of the $\chi^2$, and the sum over the $k$ modes just gives an overall factor that depends only on the details of the survey--which sets $V(z_i)$--and on the maximum wavenumber in the analysis, $k_{\rm max}(z_i)$. The latter is chosen as the minimum between the linear scale and the scale where the shot noise starts to dominate. This guarantees that our linear description is consistent and that we can safely ignore the shot-noise in $\sigma_{k,i}$. To simplify further the analysis, we will assume the same galaxy bias in every model, so that it cancels out of eq.~\eqref{eq:like}. Moreover, we will not take into account redshift space distortions, because theoretically it probes the same quantity, the linear growth, so that we do not expect differences in the final results.
\vspace{-0.5cm}
\subsection{Weak lensing}

For  weak lensing, we consider lensing tomography \cite{Hu:1999ek}. The angular cross-correlation spectra of the lensing cosmic shear for a set of galaxy redshift distributions $n_i(z)$ is given by
\be \label{lenspowspec}
P^{\rm WL}_{ij} (\ell)=  \frac{\ell}{4} \int_0^{\infty} \frac{dz}{H(z)}\, \frac{W_i(z) W_j(z) }{ \chi^3(z)}\, k_\ell^3 (z) P_{\Phi + \Psi} [ z, k_\ell(z)], 
\ee
where $\chi(z) \equiv \int_0^z dz/H (z)$ is the  comoving distance and the lensing efficiency in each bin is given by
\be
W_i (z) \equiv \chi(z) \int_z^{\infty} d \tilde z \, n_i(\tilde z)  [\chi(\tilde z) - \chi(z)]/\chi(\tilde z)  \;,
\ee
with each galaxy distribution normalized to unity, $\int_0^\infty dz \, n_i(z) = 1$.
Moreover,
$P_{\Phi + \Psi} (k)$ is the  power spectrum of  $\Phi + \Psi$. Using eq.~\eqref{LensingPot}, it is related to the matter power spectrum by
\be
P_{\Phi + \Psi} (z,k)  =   -\frac{3a^2H^2 \Omega_{\rm m}\mu_{\rm light}}{k^{2}}\,\frac{\delta_{\rm m} (z)}{\delta_{\rm m} (z=0)} P_{0} (k) \;,\label{PWL}
\ee
where $P_{0} (k)$ is the power spectrum at $z=0$.
Finally, we define $k_\ell (z) \equiv \ell /\chi(z)$ as the wavenumber which projects into the angular scale $\ell$, which we will take to vary between $[10,1000]$. Furthermore, we follow e.g.~\cite{Casas:2017eob} and assume a photometric survey of $15 \, 000$ squared degrees in the redshift range $0 < z < 2.5$, with a redshift uncertainty $\sigma_z (z) = 0.05 (1+z)$, and a galaxy distribution  $
n (z) \propto  z^2 \exp\left[ - \left( z/z_{ 0} \right)^{1.5} \right]$ \cite{Smail:1994sx}, where $z_0 = z_{m}/1.412$ and $z_{m}$ is the median redshift, assumed to be $z_{m}=0.9$ \cite{Amara:2006kp,Amendola:2012ys}.
Then, we  divide the survey into 12 equally populated redshift bins. For each bin $i$, we define the distribution $n_i(z)$ by convolving $n(z)$ with a Gaussian whose dispersion is equal to the photometric redshift uncertainty $\sigma_z(z_i) $, $z_i$ being the center of the $i$th bin (see also \cite{Giannantonio:2011ya,Amendola:2011ie}).
Adding a diagonal term to account for intrinsic ellipticity of galaxies (see e.g. \cite{Joachimi:2007xd,Takada:2008fn}), we find
\be
C_{ij} (\ell)\equiv P^{\rm WL}_{ij} (\ell)+\delta^i_j\sigma_\epsilon^2 \bar n_i^{-1}\,,\label{dejCij}
\ee
where $\bar n_i=3600*(180/\pi)^2n_\theta/N_{\rm bins}$ is the average number of galaxies per radian$^2$ per bin, assuming a total number of galaxies per arcmin$^2$  $n_\theta=30$ and $N_{\rm bins}$ equally populated bins. The intrinsic ellipticity is characterized by $\sigma_\epsilon$, which we take to be $0.22$ (EUCLID-like characteristics, see e.g.~\cite{Casas:2017eob}).
We then assume a gaussian likelihood, with covariance given by 
\be
C^{\rm true}_{ij} (\ell)\equiv P^{\rm WL,true}_{ij} (\ell)+\delta^i_j\sigma_\epsilon^2 \bar n_i^{-1}\,,
\ee
so that the $\chi^2$ is given by
\be
\begin{split}
&\chi^2_{{\rm WL},\mathcal{M}}\equiv f_{\rm sky}\sum_{\ell_{\rm min}}^{\ell_{\rm max}}(2\ell+1) \times \\
& \Tr[(C^{\mathcal{M}}_{\ell}-C^{\rm true}_{\ell})\cdot(C^{\rm true}_{\ell})^{-1}\cdot(C^{\mathcal{M}}_{\ell}-C^{\rm true}_{\ell})\cdot(C^{\rm true}_{\ell})^{-1}]\,,
\end{split}
\ee
where $f_{\rm sky}=0.36$, $l_{\rm min}=10$ and $l_{\rm max}=1000$.
Contrarily to the case of galaxy clustering, the scale dependence of the power spectrum does not factor out of the $\chi^2$. Thus, we need to fix $P_0(k)$ in eq.~\eqref{PWL}. To do so, we use CAMB \cite{Lewis:1999bs} to compute the power spectrum in $\Lambda$CDM at $z=0$, $P^{\Lambda{\rm CDM}}_0$. Then, since the scale dependence is not changed in our scenario, we have 
\be
P_0^{\mathcal{M}}(k)= P^{\Lambda{\rm CDM}}_0(k)\left(\delta_{\rm m}^{\mathcal{M}}/\delta_{\rm m}^{\Lambda{\rm CDM}}\right)^2(z=0)\,,
\ee
where $\delta_{\rm m}^{\mathcal{M}}$ is the linear growth computed in the modified gravity model $\mathcal{M}$, and $\delta_{\rm m}^{\Lambda{\rm CDM}}$ the one in $\Lambda$CDM.

 \section{The results}
 \label{sec:results}

 Even before trying to fit the parametrizations of Section \ref{sec:params}, taking many realizations of eq.~\eqref{truealpha} allows us to make general statements about the theory space itself, which we will explain in Section \ref{beforeBIC}.  Then we will go into the details of how the parametrizations perform in Section \ref{withBIC}. While for both sections, the detailed and quantitative results depend on the exact form chosen in eq.~\eqref{truealpha}, the qualitative interpretation should not change too much, provided the form of $\{\aX\}$ is general enough.

 \subsection{From theory to observables}
\label{beforeBIC}
\begin{figure}[h!]
\centering
\includegraphics[width=0.44\textwidth]{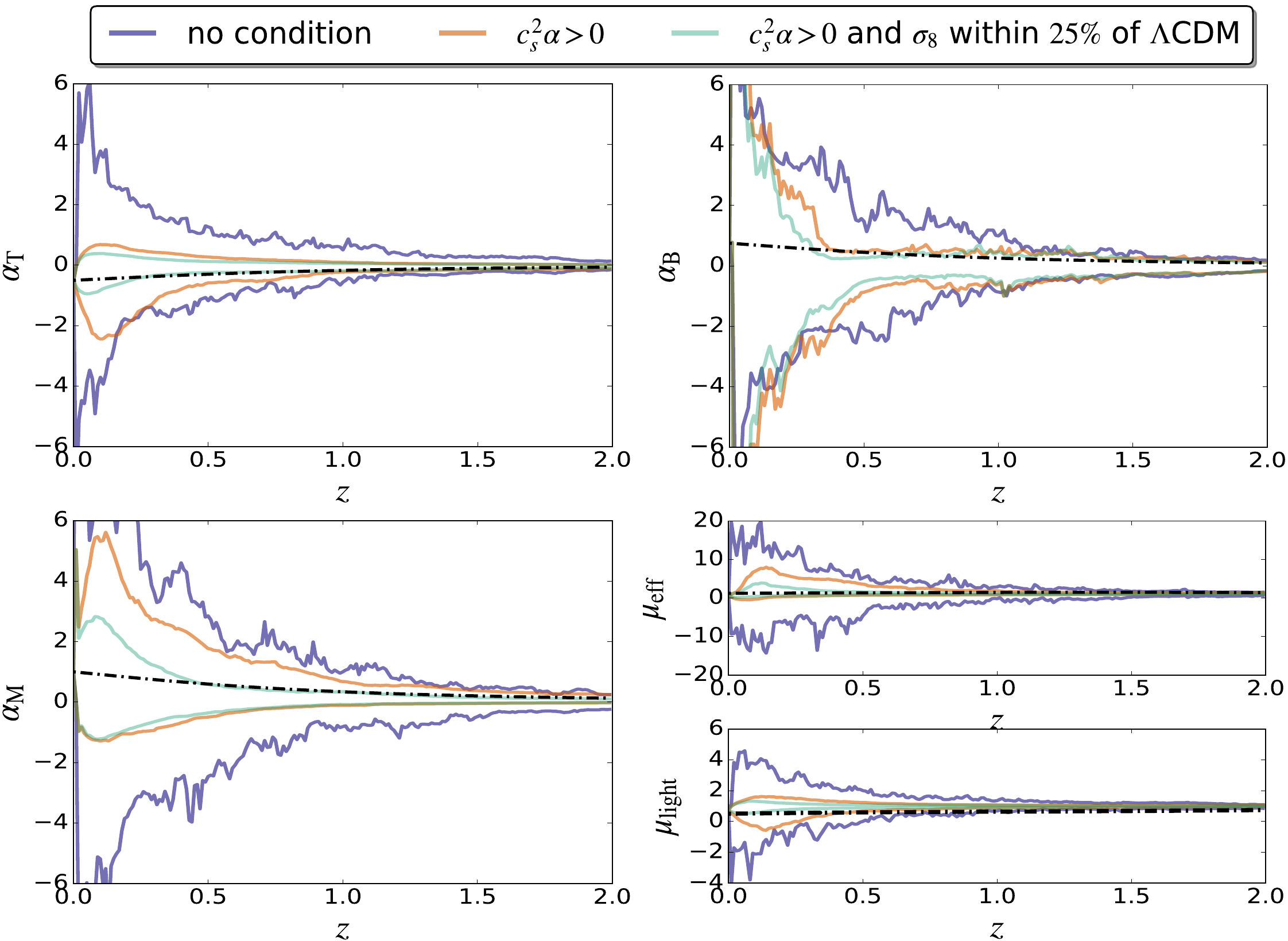}
\caption{Comparisons of the distribution of redshift evolution of $\aT$ ({\it top left}), $\aB$ ({\it top right}), $\aM$ ({\it bottom left}), $\mu_{\rm eff}$  and $\mu_{\rm light}$ ({\it bottom right}) for $\{\alpha_{{\rm B},0},\alpha_{{\rm T},0},\alpha_{{\rm M},0}\}=\{-0.5,0.75,1\}$. The lines delimit where $95\%$ of the curves reside. In dark purple, the case where the is no restriction on $c_s^2\alpha$. In orange, the parameters taken only if the corresponding $c_s^2\alpha$ is positive. Finally, in  light green, the case where $c_s^2\alpha>0$ and $\sei$, as well as $\mu_{\rm light}\sei$, are within $25\%$ of the $\Lambda$CDM value. The black dotted-dashed lines correspond to the $\{\aX\}$ evolving according to eq.~\eqref{paramOm}.}
\label{fig:combinedcs}
\end{figure}

 The first thing that should be noted is that imposing the stability condition $c_s^2 \alpha>0$ drastically reduces the size of the parameter space. We illustrate this in fig.~\ref{fig:combinedcs}, where we summarize the distribution of evolutions for the $\{\aX\}$ for $10^4$ realizations of three cases. The lines  show where $95\%$ of the curves lie. In dark purple, no conditions are imposed. In orange, we restrict realizations to $c_s^2 \alpha>0$. Only $\sim 0.2\%$ of the functions satisfy this condition with linear priors on all the parameters in eq.~\eqref{truealpha}. Finally, in light green, we restrict to $c_s^2 \alpha>0$ and to having $\sei$ and $\mu_{\rm light}\sei$ within $25\%$ of the $\Lambda$CDM value, to be consistent with past surveys. 
\vspace{-0.25cm}
\vspace{-0.25cm}

One can see that already the condition $c_s^2 \alpha>0$ reduces significantly the parameter space. It also makes the functions much smoother than the naive expectation. $\aB$ is somewhat special: it enters in a complicated, nonlinear way (it is also the only function that appears with a derivative) in $c_s^2$ and imposing $c_s^2\alpha>0$ is not as constraining as for the other parameters, see the top right panel of fig.~\ref{fig:combinedcs}. On the other hand, the same argument means that wildly varying $\aB$ do not necessarily lead to strong features in $\mu_{\rm eff}$, as seen on the bottom left panel of fig.~\ref{fig:combinedcs}.

\begin{center}

\begin{figure}[h!]
\centering
\includegraphics[width=0.93\textwidth]{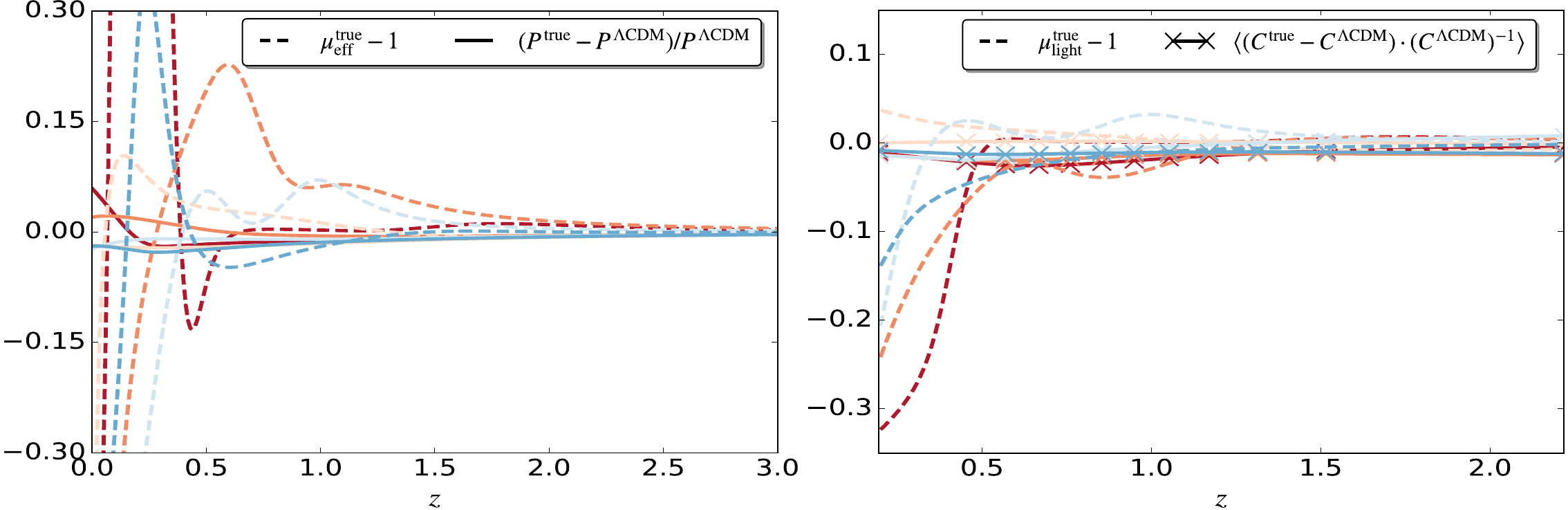}
\caption{ The relative difference with $\Lambda$CDM in $\mu_{\rm eff}$ and $\mu_{\rm light}$ (dashed) compared to the relative difference in the corresponding observables (same color, full line), respectively the galaxy power spectrum and the $C_{ij}$ averaged over $\ell$. The five colors are for 5 different random realizations of eq.~\eqref{truealpha}. For weak lensing, the crosses indicates the center of the redshift bins. While $\mu_{\rm eff}$ and $\mu_{\rm light}$ vary on short time scale and by significant amounts (more than $200\%$ for $\mu_{\rm eff}$), the observables are much smoother, and the deviations much smaller.}
\label{fig:obsosc}
\end{figure}
\end{center}

To each point $\{\alpha_{{\rm B},0},\alpha_{{\rm T},0},\alpha_{{\rm M},0}\}$ is associated $n_{\rm runs}=200$ random realizations of eq.~\eqref{truealpha} which all satisfy $c_s^2\alpha>0$ for the whole redshift range we consider. For each realization, we compute the matter growth as well as lensing potential, in order to get the galaxy and weak lensing power spectra. The second thing that we notice here is that, even if the complex evolution of the $\{\aX\}$ leads to strongly varying $\mu_{\rm eff}$ and $\mu_{\rm light}$, this is not completely transferred to the observables, as seen in fig.~\ref{fig:obsosc}. 

This is because to get the growth, one has to integrate eq.~\eqref{eq:growth}, which means variations in $\mu_{\rm eff}$ are not directly transferred to variations in the power spectrum. For weak lensing, $\mu_{\rm light}$ appears explicitly, and one might think that the effects should be more apparent. However, to get the lensing power spectrum in eq.~\eqref{lenspowspec}, one has to integrate over the window functions, which also smoothes the variations in $\mu_{\rm light}$. To estimate the deviations in the weak lensing power spectra, we use following the quantity

\be
\label{avgCl}
\begin{split}
\langle(C^{\rm true}-C^{\mathcal{M}}) \cdot (&C^{\rm true})^{-1}\rangle(z_i)\equiv \\
&\frac{\sum_\ell \left[(C_{\ell}^{\rm true}-C_{\ell}^{\mathcal{M}}) \cdot(C_{\ell}^{\rm true})^{-1}\right]_{ii}}{l_{\rm max}-l_{\rm min}} \,,
\end{split}
\ee
where $z_i$ is the middle of the redshift bin $i$.

\onecolumngrid

\twocolumngrid

Schematically, what we have explained in this section is the left half of fig.~\ref{fig:Diagram}: the full space of functions (light blue) is larger than the one of stable functions (dark blue), which in turn is larger than the space of observables (orange). To explain the other half of that figure, we will try and fit the simple parametrizations of Section \ref{sec:params} to the observables.

\subsection{Fitting the observables}
\label{withBIC}
\begin{figure}[h!]
\centering
\includegraphics[width=0.5\textwidth]{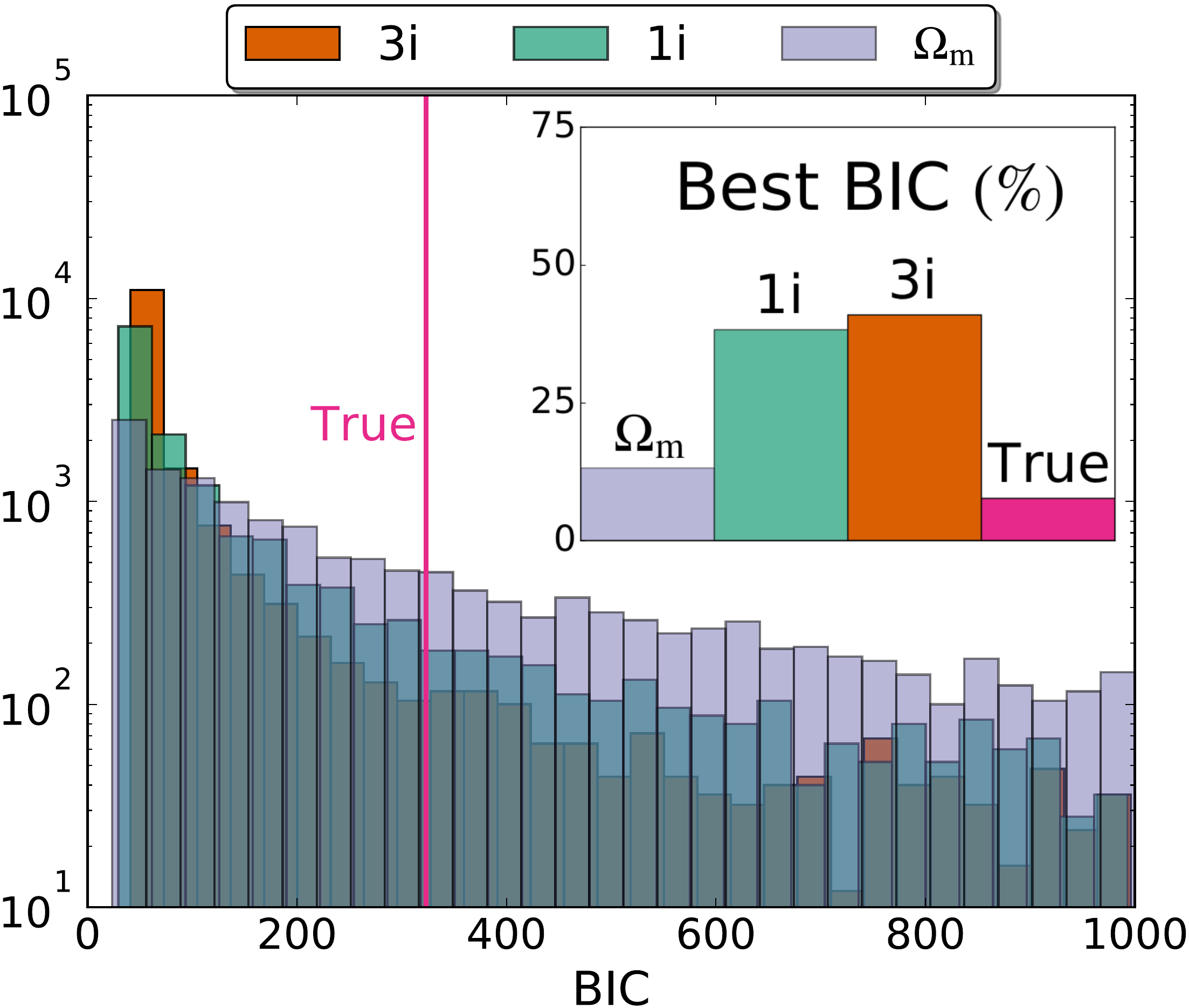}
\caption{Histogram of the BIC values for ``$\Omega_{\rm m}$" (light purple), ``1i" (green), ``3i" (dark orange), compared to the ``true" value (pink line). The last one corresponds to a perfect fit $\chi^2=0$ with eq.~\eqref{truealpha}, but is penalized in the BIC due to its high number of parameters. Most ($93\%$) of the time, at least one simpler model has a significantly lower BIC ($\Delta {\rm BIC}<-10$) than ``true", meaning it is preferred over the complex model. In the corner, we show the percentage of times when a given model has the lowest BIC. }
\label{fig:histBIC}
\end{figure}

Here, we minimize $\chi_{\mathcal{M}}^2\equiv \chi^2_{{\rm WL},\mathcal{M}}+\chi^2_{{\rm PS},\mathcal{M}}$, assuming the model $\mathcal{M}$ is given by either $w$CDM, the parametrization ``$\Omega_{\rm m}$", or the parametrizations ``1i" as well as ``3i" as detailed in Section \ref{sec:params}. From there, we can compute the BIC \eqref{defBIC}, and see which model is favored (the one with the lowest BIC). We show the results in fig.~\ref{fig:histBIC}, where we have restricted our analysis to cases where the probability of an observed $\chi^2$ being larger than $\chi^2_{{\rm min,} w{\rm CDM}}$ is smaller than $1\%$. This is to focus on cases which would lead to a detection of modified gravity, not simply a different equation of state for dark energy. 

For most realizations, the simple parametrizations do better than the full model \eqref{truealpha} in terms of BIC. Having a lower BIC does not necessarily mean being a good fit. For those with a BIC lower than the ``true" model's however, one finds that they are always a reasonable fit to the observables\footnote{We have about 360 measurement points in our analysis.}. 

Quantitatively, we find that compared to the ``true" model,  ``$\Omega_{\rm m}$", in light purple, is highly favored ($\Delta {\rm BIC}<-10$) $54\%$ of the time, for ``1i" (green) this rises to $78\%$, and to $86\%$ for ``3i" (dark orange). That is to say, the high complexities of the $\{\aX\}$ is not transferred into the observables, at least at the sensitivity level of next-generation surveys. As a consequence, simple parametrizations are able to capture the behavior of modified gravity in the galaxy and weak lensing power spectra. Note that when doing the analysis with the galaxy power spectrum only, then ``$\Omega_{\rm m}$" is doing the best job $50\%$ of the time. Adding lensing makes it harder to fit both observables with this fixed time dependence, making ``3i" more adequate. 

One could rightfully argue that ``3i" comes out as the best parametrization because of the exact form of eq.~\eqref{truealpha}, which has the same factor $(1+z)^{-q_{\rm X}}$. This is certainly true, but this adds to our argument that short time-scale variations--encoded in the rational function of eq.~\eqref{truealpha}--are not observable, only smooth behaviors.

\begin{figure}[h!]
\centering
\includegraphics[width=0.49\textwidth]{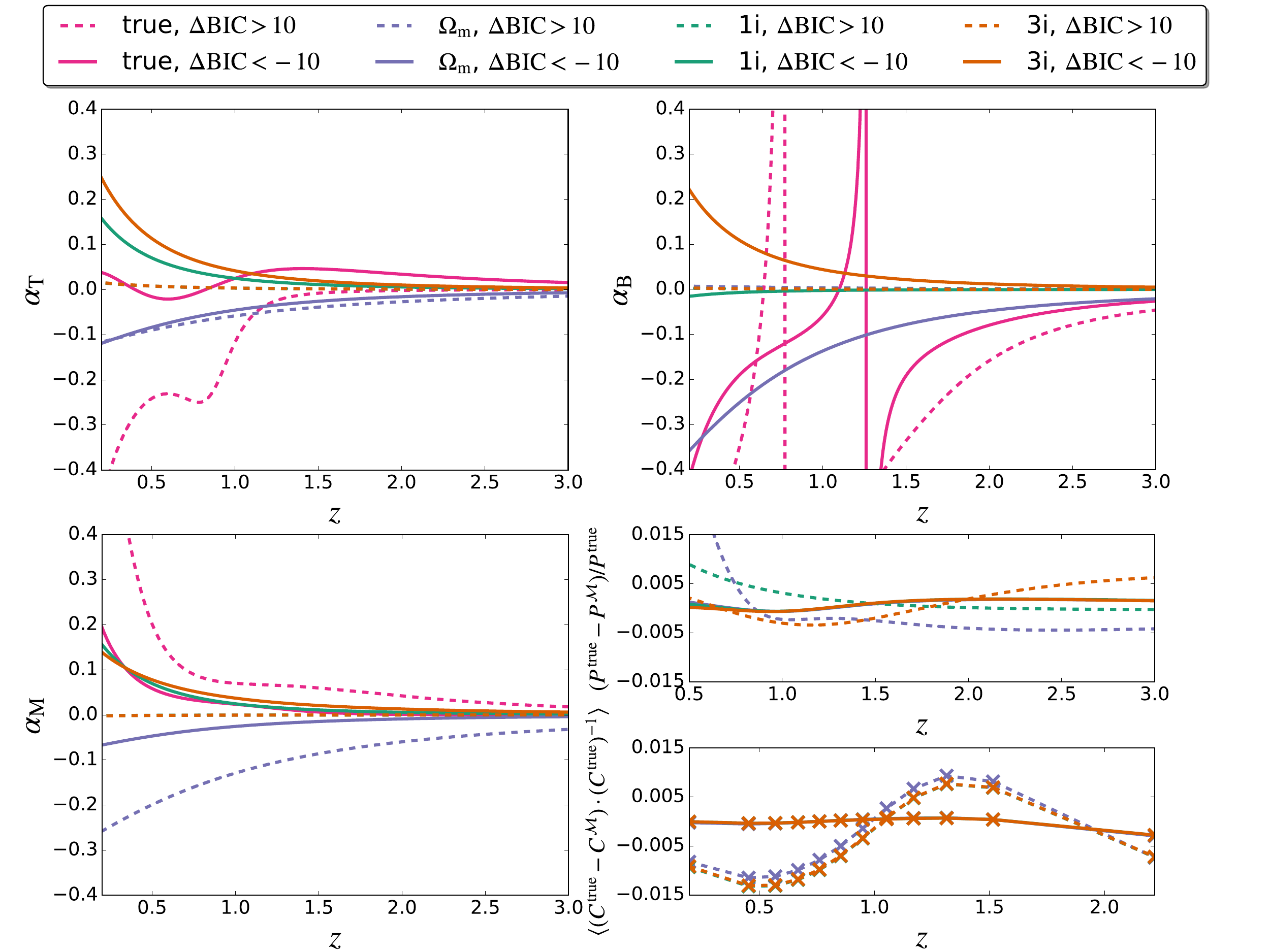}
\caption{ Evolution of the $\{\aX\}$ as a function of redshift for the ``true" model (pink), ``$\Omega_{\rm m}$" (purple), ``1i" (green) and ``3i" (orange). The full lines represent a case where the simple parametrizations are all successful at reconstructing the observables, while they all fail for the dashed lines. The bottom right panel shows the relative difference in the observables between the simple models and the ``true" ones. The crosses indicates the center of the redshift bins in the weak lensing measurement.}
\label{fig:CombinedBIC}
\end{figure}

We show in fig.~\ref{fig:CombinedBIC} the evolution of the $\{\aX\}$ as well as the relative difference in the power spectrum\footnote{This is shown as a function of $z$ and not $k$ because the scale dependence is not modified in our approach.} and $C_{\ell}$ (given in eq.~\eqref{avgCl}) for the different models $\mathcal{M}$, in two cases. With dashed lines, a case where the simple parametrizations fail at reproducing the observables (all the $\Delta {\rm BIC}_{\mathcal{M}-{\rm true}}>10$), and a case where they succeed (all the $\Delta {\rm BIC}_{\mathcal{M}-{\rm true}}<-10$) with full lines.

As in fig.~\ref{fig:obsosc}, the strong features in the $\{\aX\}$ are much smoother in the observables. In the successful case, where $\aT$ exhibits sizeable variations (top left panel), the observables are well reproduced by the smooth $\{\aX\}$ of the simple parametrizations (bottom right panel). As we have noted after fig.~\ref{fig:combinedcs}, the effect of $\aB$ on the observables is not as simple as for the other functions, meaning that very complicated evolution such as the ones on the top right panel of fig.~\ref{fig:CombinedBIC} are not problematic for fitting simple parametrizations to observables. In the successful case for $\aM$, note that indeed, the ``1i" and ``3i" capture quite well its general behavior, but miss the short time-scale variations. For the bottom right panels, the three curves for the successful case are on top of each other, and have a relative difference with the ``true" observables that is less $0.3\%$.

\section{Discussion $\&$ Conclusions}
\label{sec:ccl}
The possibility of an additional degree of freedom in gravity opens a vast theory space, described conveniently at the level of linear perturbations within the framework of the EFT of DE \cite{Gleyzes:2014rba}. Formally, this theory space is infinite, as it depends on free functions of time $\{\aX(z)\}$, not parameters. Therefore, it is not clear at first sight how one can use the EFT of DE to constrain modified gravity in upcoming surveys. The question we try to answer in this article is whether to describe observables, this infinite dimensional space can be approximated by a finite one, so that the cosmological analyses can be reduced to actual parameters, not functions.

One approach that has been used in previous studies was to explore specific models, where the $\{\aX\}$ are derived from the evolution of the background field and the parameters of the model. Their evolution is thus known, and one can try and see whether they can be approximated by simple functions, that one could then use as templates for analyzing data. While in principle this is certainly a promising way to go, we see it as limited in the context of modified gravity. The main reason is that, up to now, these studies have only focused on a small number of models, and a small number of parameters within those models, leading to contradicting results \cite{Renk:2016olm,Linder:2016wqw}. The problem is that, since that are no outstanding candidates for modified gravity, there is no reason to choose one particular model over another. Therefore, one should explore the whole model space (e.g. that of Horndeski theories \cite{Horndeski:1974wa})\footnote{This would be a good first step, but as recent developments have shown \cite{Deffayet:2015qwa,Langlois:2015cwa,Langlois:2015skt,Achour:2016rkg,Crisostomi:2016tcp}, it is not clear Horndeski is the full picture of scalar-tensor theories.}. This is similar to what has been done at the level of the background expansion in \cite{Marsh:2014xoa,Raveri:2017qvt}. However, this approach does not take full advantage of the powerful potential of EFT of DE to explore the large theory space systematically.

Thus, rather than explore the models, we take a more agnostic approach, and directly explore the $\{\aX\}$. To do so, we choose a convoluted form for their time dependence, eq.~\eqref{truealpha}, which allows a vast range of different evolutions. We have checked it can reproduce known models such as the $k$-essence \cite{ArmendarizPicon:2000dh} and galileons \cite{Deffayet:2009wt} to better than $10^{-5}$ accuracy.

 The first thing that we notice is that the function space is actually smaller than expected, because the $\{\aX\}$ must obey stability conditions (namely, positive kinetic term and positive sound speed). This drastically reduces the number of possible functions, by about a factor of 5000 (with our uniform priors on the parameters of eq.~\eqref{truealpha}). This can be visualized in fig.~\ref{fig:combinedcs}. If we then propagate the complex evolution of the $\{\aX\}$ to the matter growth and lensing, one can see in fig.~\ref{fig:obsosc} that strong variations in the $\{\aX\}$ lead to strong variations in the modified gravitational couplings $\mu_{\rm eff}$ and $\mu_{\rm light}$, but that those variations are damped when looking at observables. This is one of the key points of this paper:  it seems difficult to have access to short time variations of the $\{\aX\}$, because observables are not very sensitive to them. Going back to the schematic representation of fig.~\ref{fig:Diagram}, this shows that the orange circle is indeed smaller than the dark blue one. Therefore, we anticipate that it is legitimate to use simple parametrizations to look for deviation to GR, even though they are much smoother than one might expect from models.

This is what we explored in the rest of the article. We compared how three differents parametrizations could reproduce observables coming from a theory with complicated $\{\aX\}$. In the first one, dubbed ``$\Omega_{\rm m}$", all the $\{\aX\}$ are proportional to $1-\Omega_{\rm m}$, a common choice in the literature. For the second one, referred to as ``1i", all of the $\{\aX\}$ are proportional to the same function $(1+z)^{-q}$, where $q$ is allowed to vary between $[0,6]$. Finally, the last one, called ``3i", allows for more freedom, each $\{\aX\}$ being proportional to $(1+z)^{-q_{\rm X}}$, with a different index $\in [0,6]$ for each function. The functions $\{\aX\}$ with those parametrizations do not exhibit sharp features, but since those are not observable, they can still be used to fit the galaxy and weak lensing power spectra rather well. In a sense, the complexity in eq.~\eqref{truealpha} is not demanded by the observables, because we cannot have access to it, and therefore is not relevant (at least naively, with the sensitivity of future experiments). To classify the performance of models while penalizing unnecessary complexity, we choose to use the Bayesian Information Criterion (BIC)\footnote{One could also use the Bayesian evidence, but it is much more computationally expensive. Looking at the smooth curves in fig.~\ref{fig:obsosc}, we would expect similar results.}.

The results are shown in fig.~\ref{fig:histBIC}.  Each parametrization is compared to the ``true" case, i.e.~the original model governed by eq.~\eqref{truealpha}, which fits perfectly the data, but introduces 18 parameters for each $\{\aX\}$. In $93\%$ of the case, one of the three simple parametrizations has a lower BIC, meaning the data does not need the introduction of the full complexity of \eqref{truealpha}. Our results show that even though very high complexity is not necessary, taking all of the $\{\aX\}$ with the same time dependence might be too simplistic. The model that performed best is the one with 2 parameters to describe each $\{\aX\}$: an amplitude and power law index. 

The quantitative results of fig.~\ref{fig:histBIC} are dependent on the exact form of eq.~\eqref{truealpha} and the choice of the priors on the parameters. However, the fact that short time-scale variations of the $\{\aX\}$ are not observables is more robust. This is also the case when using a rational function of $N=-\log[1+z]$ instead of $z$ in eq.~\eqref{truealpha} while increasing the prior on $\{d_{{\rm X},i},n_{{\rm X},i}\}$ to $[-10^4,10^4]$. This can reproduce known models to better than $10^{-5}$ as well, and the qualitative behavior described in Section \ref{beforeBIC} is very similar. There are some differences, for example the number of stable functions represents $\sim 0.4\%$ of all functions instead $\sim 0.2\%$ in Section \ref{beforeBIC}. The differences in the actual numbers of Section \ref{withBIC} (e.g. those of fig.~\ref{fig:histBIC}) are more minor: ``$\Omega_{\rm m}$" does better than ``true" $55\%$ of the time, instead of $54\%$, $83\%$ instead of $78\%$ for ``1i" and $90\%$ instead of $86\%$ for ``3i".

We have focused here on how well parametrizations can fit complex models, and we did not say much about actually constraining the $\{\aX\}$. Indeed, with our simple approach in the Extreme Quasi-Static limit, the degeneracies are too strong to be broken with only galaxy clustering and weak lensing, as noted already in \cite{Gleyzes:2015rua}. Therefore, the constraints would not really be meaningful. To get a sense however, one can look at a simplified case where $\aM$ is fixed to zero. The degeneracies are not as strong in this case, and one can use Monte Carlo Marko Chains (with only $\{\alpha_{\rm X,0},q_{\rm X}\}$ as free parameters) to forecast the constraints, assuming a $w$CDM fiducial with $w=-0.95$. With ``$\Omega_{\rm m}$'', we get $\sigma(\aX)\sim 0.03$. With ``1i", this number is about 10 times larger, and ``3i" (which is rather ``2i" here) gives results comparable to ``1i". Thus, the constraints do degrade, but not by many orders of magnitude. It would be interesting to see how these numbers change with a more comprehensive analysis, such as the one with Hi-Class \cite{Zumalacarregui:2016pph}. Indeed, this code uses the full equations (no quasistatic approximation), so that Ref.~\cite{Alonso:2016suf} were able to put constraints on all the $\{\aX\}$. They used a parametrization similar to ``$\Omega_{\rm m}$" (but with an additional constant parameter) as well as more complicated time dependences. 

The optimal way to parametrize the functional freedom of the EFT of DE remains to be determined. We have shown that using simple parametrizations should do an adequate job. Another way to go could be to assign a different value to the $\{\aX\}$ in each redshift bins. However in that case, it is not clear how to enforce the stability conditions. Moreover, this would bring a very large number of parameters, which would considerably weakens the constraints, although one could use a PCA approach to extract the most constrained directions in the parameter space (see e.g. \cite{Casas:2017eob}). The advantage of the three parametrizations that we explored here is that they are simple, and somewhat physically motivated: the effects go to zero in matter domination. Plenty of other functions satisfy those two criteria, and we leave for future work a more comprehensive investigation.

\emph{Acknowledgements:}
It is a pleasure to thank Alexandre Barreira, Phil Bull, Olivier Dor\'e, Filippo Vernizzi and Miguel Zumalac\'arregui for fruitful discussions and useful comments. Part of the research described in this paper was carried out at the Jet Propulsion Laboratory, California Institute of Technology, under a contract with the National Aeronautics and Space Administration. This research is partially supported by NASA ROSES ATP 14-ATP14-0093 grant.

\bibliographystyle{utphys}
\bibliography{refs}

\end{document}